\documentclass[aps,prl,preprint]{revtex4-2}

\usepackage{amssymb}
\usepackage{booktabs}
\usepackage{graphicx}
\usepackage[version=4]{mhchem}
\usepackage{placeins}

\begin{document}

\title{Derivative Discontinuity in Many-Body Perturbation Theory and Chemical Potentials in Random Phase Approximation}

\author{Jiachen Li}
\affiliation{Department of Chemistry, Duke University, Durham, NC 27708, USA}
\affiliation{Department of Chemistry, Yale University, New Haven, CT 06520, USA}
\author{Weitao Yang}
\email{weitao.yang@duke.edu}
\affiliation{Department of Chemistry and Department of Physics, Duke University, Durham, NC 27708, USA}

\begin{abstract}
We derive analytical expressions for chemical potentials within the random phase approximation (RPA), equivalently the $GW$ energy functional evaluated using non interacting Green’s functions ($G_s$). The chemical potential is obtained using two formally equivalent approaches: a direct derivative of the total energy with respect to particle number, and a functional derivative via the chain rule through $G_s$, both validated with finite difference benchmarks. We show that the functional derivative of the $GW$ correlation energy—i.e., the $GW$ correlation self energy—exhibits a discontinuity at integer particle numbers with finite jumps. This resolves the apparent inconsistency between accurate $GW$ quasiparticle energies and the large delocalization errors observed in RPA total energies, as standard $GW$ self energies neglect this nonanalytic behavior. Our results suggest that derivative discontinuities are a fundamental feature of correlation energy functionals, analogous to the known discontinuity in the exact exchange correlation energy.
\end{abstract}

\maketitle

\newpage
Chemical potential, defined as the derivative of the total energy with the electron number, 
is a central concept for studying electronic properties in chemistry and materials science. 
It quantifies the electron addition and removal processes,
as measured by the ionization potential (IP) and the electron affinity (EA) in photoemission and inverse-photoemission spectroscopy. 
For bulk materials, the difference between the chemical potentials is the fundamental gap. 
The chemical potential plays a vital role in determining electron distribution at material interfaces.

Traditionally, the chemical potential is defined in the thermodynamic limit. 
It has been extended to finite systems in density functional theory (DFT) \cite{hohenbergInhomogeneousElectronGas1964,kohnSelfConsistentEquationsIncluding1965,parrDensityFunctionalTheoryAtoms1989} by defining fractional particle numbers for finite electron numbers in a grand canonical ensemble at zero temperature\cite{perdewDensityFunctionalTheoryFractional1982}. 
The exact energy for fractional particle numbers is a linear interpolation between the two corresponding nearest integer systems. 
The same conclusion was established from a consideration of molecular dissociation based on the superposition principle for degenerate states and the size consistency requirement \cite{yangDegenerateGroundStates2000}. 
For a physical system with an integer electron number, 
there are two chemical potentials associated with the slopes of adding and removing an electron. 
They are equal to the negative IP and EA, respectively, for the exact functional, 
in DFT or in many-electron perturbation theory (MBPT). 

How to determine chemical potentials from common (generalized) Kohn-Sham calculations was first established in 2008 by Cohen, Mori-Sanchez and Yang\cite{cohenFractionalChargePerspective2008}. 
For any (generalized) Kohn-Sham calculation, 
the chemical potential for electron removal is just the energy of  highest occupied molecular orbital (HOMO), 
and the chemical potential for electron addition is the energy of lowest unoccupied molecular orbital (LUMO) energy, 
with any continuous approximate functional of density or the density matrix\cite{cohenFractionalChargePerspective2008}--this is the ground state chemical potential theorem. It goes beyond the meaning of orbital energies in the Janak theorem\cite{janakProofThatFrac1978}. 
This validates the use of HOMO and LUMO energy to approximate IP and EA for molecules and the fundamental gap for materials, based on the meaning of exact chemical potentials.

Although it is a valid approximation, 
the accuracy of the approximation to IP and EA from the HOMO and LUMO energy depends on the quality of the approximate functional used. 
It is well-known that DFT with commonly used approximations, 
including the local density approximation (LDA), generalized gradient approximations (GGA) and hybrid functionals, 
systematically underestimate IP and overestimate EA for molecules, 
similarly underestimate the band gap of solids\cite{perdewPhysicalContentExact1983, shamDensityFunctionalTheoryEnergy1983,cohenChallengesDensityFunctional2012,perdewDensityfunctionalEnergyGaps2018} and lead to energy level misalignment at interfaces\cite{liuEnergyLevelAlignment2017,meiSelfConsistentCalculationLocalized2020,liuDensityFunctionalDescriptions2023}. 
The root of the issue is the delocalization error of the approximate energy functionals\cite{mori-sanchezLocalizationDelocalizationErrors2008}. 
The delocalization error exhibits a size-dependent behavior\cite{mori-sanchezLocalizationDelocalizationErrors2008,yangDerivativeDiscontinuityBandgap2012,meiSelfConsistentCalculationLocalized2020}: 
(1) For small systems with few atoms and limited physical extent, 
commonly used density functional approximations (DFAs) provide accurate descriptions of total energies for systems with integer numbers\cite{furcheDevelopingRandomPhase2008,eshuisElectronCorrelationMethods2012}. 
However, the delocalization error manifests as a convex deviation from the Perdew-Parr-Levy-Balduz (PPLB) linearity condition\cite{perdewDensityFunctionalTheoryFractional1982} for systems with fractional electron numbers. 
(2) For large systems with many atoms,  
DFAs exhibit small errors for systems with fractional electron numbers and satisfy the fractional charge linearity condition in the bulk limit. 
However, the delocalization error manifests as significant inaccuracies in the total energies of systems with integer electron numbers, 
mainly for the charged states. 
(3) For finite systems with large physical extent (e.g., near dissociation limits), 
the delocalization error introduces inaccuracies in the total energies of systems with integer electron numbers, 
in addition to the convex deviation from the linearity condition for fractional electron numbers. 

The derivative discontinuity of the energy functional plays a key role in understanding the approximations. 
LDA and GGA exchange correlation energies are continuous functionals of the electron density with no derivative discontinuity,
and the HOMO and LUMO energies are their corresponding approximations to the negative of IP and EA. 
Due to the delocalization error, both LDA and GGA provide poor estimate of IP and EA through the HOMO and LUMO energies. 
This can also be viewed as the lack of the derivative discontinuity, 
because the exact exchange and correlation energy functional cannot be a continuous functional of electron density\cite{perdewPhysicalContentExact1983, shamDensityFunctionalTheoryEnergy1983,cohenChallengesDensityFunctional2012}. 
Hybrid functionals are continuous functionals of the noninteracting density matrix, 
but not of the density. 
Their nonlocal generalized Kohn-Sham potentials lead to improved HOMO and LUMO energies for predicting IP and EA
And the corresponding local potentials, defined as the functional derivative with respect to the electron density, 
would have derivative discontinuity at integer number of electron\cite{cohenFractionalChargePerspective2008,cohenChallengesDensityFunctional2012,yangDerivativeDiscontinuityBandgap2012}. 
Based on the exact conditions for fractional charges and spins combined, 
Mori-Sanchez, Cohen and Yang have shown that the exact exchange-correlation energy functional cannot be a continuous and differential functional of the non-interacting density matrix\cite{mori-sanchezDiscontinuousNatureExchangeCorrelation2009}. 
It highlights the importance of explicitly discontinuity in the density matrix for describing strongly correlated systems.

Beyond hybrid functionals,
MBPT has also been widely used to model electronic properties of molecular and periodic systems, both in and outside DFT.
In MBPT approaches,
the quasiparticle energies as Dyson orbital energies, equal to the total energy differences at integer electron numbers play a key role in predicting IP and EA\cite{hedinNewMethodCalculating1965,martinInteractingElectrons2016,bruneval$GW$ApproximationManyBody2009,brunevalIonizationEnergyAtoms2012}. 
The basic variable in MBPT is the one-particle Green's function $G$.
Just as the Hartree-Fock (HF) theory can be viewed as a DFA, 
MBPT can lead to approximate energy functionals in the context of DFT \cite{grabowskiInitioDensityFunctional2007,cohenSecondOrderPerturbationTheory2009, grabowskiComparingInitioDensityfunctional2011} when we use non-interacting Green's function $G_s$ as the basic variable, 
which makes the GW approximate correction energy into the random phase approximation (RPA)\cite{martinInteractingElectrons2016}. 

RPA was originally designed to describe the screening effect in the electron gas\cite{bohmCollectiveDescriptionElectron1953,gell-mannCorrelationEnergyElectron1957} ranks among the most popular approaches.
RPA can be equivalently derived from different approaches,
including Hedin's equations in the Green's function theory\cite{hedinNewMethodCalculating1965,dahlenVariationalEnergyFunctionals2005,martinInteractingElectrons2016,reiningGWApproximationContent2018},
the adiabatic connection fluctuation theorem\cite{furcheMolecularTestsRandom2001,furcheDevelopingRandomPhase2008,renRandomphaseApproximationIts2012,eshuisElectronCorrelationMethods2012,chenRandomPhaseApproximationMethods2017},
and the equation of motion\cite{scuseriaGroundStateCorrelation2008,berkelbachCommunicationRandomphaseApproximation2018}.
Because of the inclusion of the long-range screening,
RPA is adequate for describing nonlocal electron correlation in molecules and solids,
which shows promising results for atomization energies\cite{ruzsinszkyRPAAtomizationEnergy2010}, 
surface absorption energies\cite{renExploringRandomPhase2009,lauQuantumPlasmonsIntraband2020}, 
formation energies\cite{brunevalRangeSeparatedApproachRPA2012},
forces (analytic gradients)\cite{burowAnalyticalFirstOrderMolecular2014,rambergerAnalyticInteratomicForces2017,tahirLocalizedResolutionIdentity2022},
ionization potentials and electron affinities\cite{brunevalIonizationEnergyAtoms2012,hellgrenStaticCorrelationElectron2015}.
Combining with localized orbital techniques\cite{shiLibRPASoftwarePackage2025,liangEfficientImplementationRandom2024} and optimal grids for numerical integrations\cite{,eshuisFastComputationMolecular2010,kaltakLowScalingAlgorithms2014},
RPA becomes a low-scaling and reliable approach for describing large-scale and extended systems\cite{batesReferenceDeterminantDependence2017,ruzsinszkyInsightOrganicReactions2015,olsenRPAGWMethods2019}.

Even though MBPT has been founded on grand canonical ensemble theory,
traditionally MBPT has been developed for physical systems with only integer number electrons\cite{stefanucciNonequilibriumManyBodyTheory2013,martinInteractingElectrons2016}. 
The fractional formulation of MBPT approaches for a given approximation has been established by Yang, Mori-Sanchez and Cohen  based on the ensemble of $G_s$\cite{yangExtensionManybodyTheory2013}, 
and has been successfully applied to reveal the extent of delocalization error in MBPT approaches including second-order perturbation theory (MP2)\cite{cohenSecondOrderPerturbationTheory2009,suIntegrationApproachSecondOrder2015,liChemicalPotentialsOneElectron2024}, 
RPAs\cite{mori-sanchezFailureRandomPhase2009,mori-sanchezFailureRandomphaseapproximationCorrelation2012}
and particle-particle random phase approximation (ppRPA)\cite{vanaggelenExchangecorrelationEnergyPairing2013}.

Within the fractional formulation\cite{yangExtensionManybodyTheory2013},
the MP2 chemical potential can be equivalently obtained from the analytical derivative approach\cite{cohenSecondOrderPerturbationTheory2009,suIntegrationApproachSecondOrder2015} that directly evaluates the derivative of the MP2 energy to the occupation number
or from the functional derivative approach that employs the chain rule via $G_s$\cite{liChemicalPotentialsOneElectron2024}. 
It shows that the MP2 chemical potential is similar to the frontier quasiparticle energy from the second-order Green's function theory (GF2) and provides error around 0.5 eV for predicting IPs and EAs for molecules\cite{renResolutionofidentityApproachHartree2012,suIntegrationApproachSecondOrder2015,liChemicalPotentialsOneElectron2024}. 

\begin{figure}
\includegraphics[width=0.7\textwidth]{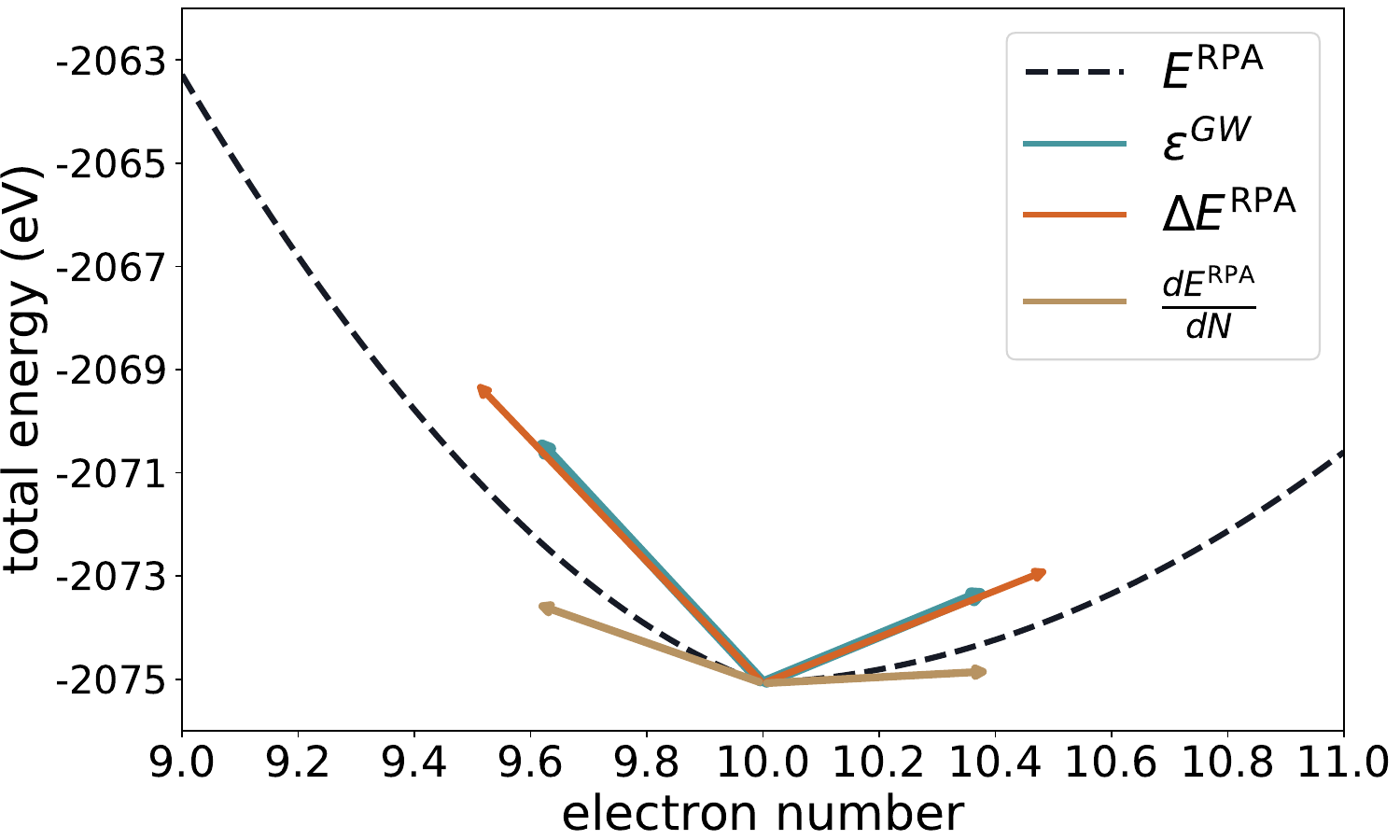}
\caption{Behavior of the RPA energy of \ce{H2O} as a function of the electron number.
Arrows represent ionization potentials and electron affinities calculated by $GW$ quasiparticle energies, $\Delta$RPA
and derivatives of the RPA energy to the particle number.
HF was used as the mean-field reference.
cc-pVDZ basis set was used.} 
\label{fig:h2o}
\end{figure}

Unlike the case of MP2,
the RPA chemical potentials remain a puzzle up to now.
The Klein functional connects the RPA correlation energy to the $GW$ correlation energy\cite{martinInteractingElectrons2016},
and the RPA (or $GW$) self-energy has been taken as the derivative of the RPA correlation energy functional with respect to $G_s$\cite{wangRelationshipRandomphaseApproximation2010,vooraVariationalGeneralizedKohnSham2019,vooraEffectiveOneparticleEnergies2019}. 
In addition, the $GW$ quasiparticle energies were claimed to be the chemical potentials, 
that is derivatives of the RPA total energy to the particle number\cite{wangRelationshipRandomphaseApproximation2010}. 
Since the $GW$ self-energy leads to quasiparticle energies that are good approximations to IP and EA for molecules and band gaps for solids, 
this would indicate that the RPA energy functional should have minimal delocalization error. 
This is contrary to the observation that RPA has large delocalization error,
shown as in fractional charge  behavior and also in molecular ion bond dissociations\cite{mori-sanchezFailureRandomPhase2009,mori-sanchezFailureRandomphaseapproximationCorrelation2012}. 
The delocalization error in RPA is much larger than that of LDA and GGA, 
which indicates massive error of using the RPA chemical potentials for predicting IPs and EAs. 
As demonstrated in Fig.\ref{fig:h2o}, 
IPs and EAs of the water molecule calculated by $GW$ and $\Delta$RPA are similar,
which are substantially different from the RPA energy derivatives.
These are contradictory results on the delocalization error in the RPA energy and its chemical potentials, 
which has not been understood.  
Why do the RPA chemical potentials differ from the $GW$ quasiparticle energies?  What is wrong with the derivation of chemical potentials through the functional derivatives with respect to the Green's function?

In this letter,
we examine the RPA chemical potentials with two approaches.
We first derive the analytical expression for the RPA chemical potentials by directly taking the derivative of the RPA correlation energy with respect to the particle number via the frontier occupation numbers\cite{cohenFractionalChargePerspective2008},
which is denoted as the direct derivative approach.
Then because RPA correlation energy can be expressed as a functional of $G_s$,
we follow our recent work for the MP2 theory \cite{liChemicalPotentialsOneElectron2024},
using the functional derivative approach to calculate the derivative of the RPA correlation energy with respect to the occupation number via the chain rule with $G_s$.
Unlike the results for the MP2 chemical potentials \cite{liChemicalPotentialsOneElectron2024},
the functional derivative approach using the RPA self-energy evaluated at integer electron numbers fails to provide the RPA chemical potentials, 
as compared to numerical finite-difference results.
We demonstrate that the agreement to numerical finite-difference results can only be obtained by evaluating the RPA self-energy at small fractional electron numbers,
which clearly reveals that the $GW$ correlation energy as a functional of Green's function has derivative discontinuity at integer number of electrons. 

In the framework of DFT, 
the RPA correlation energy can be derived from the adiabatic connection\cite{hesselmannRandomphaseApproximationCorrelation2011,eshuisElectronCorrelationMethods2012,engelDensityFunctionalTheory2011,chenRandomPhaseApproximationMethods2017}.
The matrix representation of the RPA problem extended to systems with fractional charges\cite{mori-sanchezFailureRandomPhase2009,mori-sanchezFailureRandomphaseapproximationCorrelation2012}
\begin{equation}\label{eq:rpa_sym}
    \begin{pmatrix}
        \mathbf{A}^{\text{sym}}  & \mathbf{B}^{\text{sym}} \\
        -\mathbf{B}^{\text{sym}} & -\mathbf{A}^{\text{sym}}
    \end{pmatrix}
    \begin{pmatrix}
        \mathbf{\tilde{X}} \\
        \mathbf{\tilde{Y}}
    \end{pmatrix}
    = \Omega
    \begin{pmatrix}
        \mathbf{\tilde{X}} \\
        \mathbf{\tilde{Y}}
    \end{pmatrix}\text{,}
\end{equation}
where in the symmetric formulation\cite{yangExtensionManybodyTheory2013}
\begin{align}
    A^{\text{sym}}_{ia,jb} = & \delta_{ij} \delta_{ab} (\epsilon_a - \epsilon_i) 
    + \sqrt{n_i n_j (1-n_a) (1-n_b)} \langle ib | aj \rangle  \label{eq:rpa_sym_a} \text{,}\\
    B^{\text{sym}}_{ia,jb} = & \sqrt{n_i n_j (1-n_a) (1-n_b)} \langle ij | ab \rangle \label{eq:rpa_sym_b} \text{.}
\end{align}
Here $n$ is the occupation number, 
$\Omega$ is the excitation energy, 
$\epsilon$ is the orbital energy and two-electron integral is defined as
$\langle pq | rs \rangle = \int dx dx' \frac{\psi^*_p(x) \psi_r(x) \psi^*_q(x') \psi_s(x')}{|r-r'|}$.
With the transition amplitudes $X^m_{ia} = \langle \Psi_0 | \hat{a}^\dagger_i \hat{a}_a | \Psi_m \rangle$ and $Y^m_{ia} = \langle \Psi_m | \hat{a}^\dagger_i \hat{a}_a | \Psi_0 \rangle$,
the fractional-transformed eigenvectors of the symmetric RPA matrix are defined as $\tilde{X}^m_{ia}=\frac{X^m_{ia}}{\sqrt{(1-n_a)n_i}}$ and $\tilde{Y}^m_{ia}=\frac{Y^m_{ia}}{\sqrt{(1-n_a)n_i}}$.
We use $i$, $j$ for occupied orbitals,
$a$, $b$ for virtual orbitals, 
$f$ for the fractionally occupied orbital,
$p$, $q$ for general orbitals and $m$ for excitations.
Note that the fractionally occupied orbital is considered twice as occupied orbital and also virtual\cite{yangExtensionManybodyTheory2013,mori-sanchezFailureRandomPhase2009,mori-sanchezFailureRandomphaseapproximationCorrelation2012}.

Equivalently,
to study the infinitesimal fraction limits, it is more convenient to formulate the RPA matrix in an equivalent asymmetric manner
\begin{equation}\label{eq:rpa_asym}
    \begin{pmatrix}
        \mathbf{A}^{\text{asym}}  & \mathbf{B}^{\text{asym}} \\
        -\mathbf{B}^{\text{asym}} & -\mathbf{A}^{\text{asym}}
    \end{pmatrix}
    \begin{pmatrix}
        \mathbf{X} \\
        \mathbf{Y}
    \end{pmatrix}
    = \Omega
    \begin{pmatrix}
        \mathbf{X} \\
        \mathbf{Y}
    \end{pmatrix}\text{,}
\end{equation}
where
\begin{align}
    A^{\text{asym}}_{ia,jb} = & \delta_{ij} \delta_{ab} (\epsilon_a - \epsilon_i) 
    + n_i (1-n_a) \langle ib | aj \rangle \text{,} \label{eq:rpa_asym_a} \\
    B^{\text{asym}}_{ia,jb} = & n_i (1-n_a) \langle ij | ab \rangle \text{.}\label{eq:rpa_asym_b} 
\end{align}
$\mathbf{X}$ and $\mathbf{Y}$ are the right eigenvectors of the asymmetric RPA matrix.
Because the asymmetric RPA matrix is non-Hermitian,
its left and right eigenvectors are generally different.
The derivation of symmetric and asymmetric formulations of the RPA equation is shown in Section.1 in the Supplemental Material.
The RPA equation in Eq.\ref{eq:rpa_sym} and Eq.\ref{eq:rpa_asym} for systems with a fractional electron number has the same structure as that for systems with an integer electron number,
but with with different dimensions, 
depending on the sign of the fractional electron number.
In the ($N+\delta$)-electron system,
we use explicitly the fractional $\delta$ to indicate the dependence on the limiting process,
with $\delta<0$ for the electron removal and $\delta>0$ for electron addition.
Because the fractionally occupied orbital is included in both occupied and virtual orbitals\cite{yangExtensionManybodyTheory2013},
the RPA eigenvectors have $N_{\text{occ}}$ extra dimensions for $\delta<0$ and $N_{\text{vir}}$ extra dimensions for $\delta>0$,
where $N_{\text{occ}}$ and $N_{\text{vir}}$ is the number of occupied and virtual orbitals of the $N$-electron system.
As shown in the Supplementary Material, 
RPA eigenvectors of infinitesimal fractional systems can be constructed from the integer-electron system.

As shown in Ref.\citenum{cohenFractionalChargePerspective2008},
the correlation part of the chemical potential can be obtained as the derivative of the correlation energy $E_{\text{c}}$ to the occupation number of the frontier orbital $n_f$
\begin{equation}\label{eq:rpa_mu}
    \mu_{\text{c}}
    = \frac{d E_{\text{c}}}{d N } \bigg |_{\pm}
    = \frac{d E_{\text{c}}}{d n_f } \bigg |_{\pm}\text{,}
\end{equation}
where $\pm$ stands for the left and right derivative.

We first take the direct derivative approach for the RPA chemical potentials in Eq.\ref{eq:rpa_mu} using the asymmetric RPA equation.
With the excitation energy $\Omega$ from solving Eq.\ref{eq:rpa_asym},
the RPA correlation energy are given by\cite{mori-sanchezFailureRandomPhase2009,mori-sanchezFailureRandomphaseapproximationCorrelation2012,yangExtensionManybodyTheory2013}
\begin{align}
    E_{\text{c}}^{\text{RPA}} = & \frac{1}{2} \sum_m \Omega_m - \text{Tr} \mathbf{A}^{\text{asym}}\text{.}
\end{align}
Then the correlation part of the RPA chemical potential is
\begin{equation}\label{eq:rpa_derivative}
    \mu^{\text{RPA}}_{\text{c}} = \frac{1}{2}
    \sum_m \frac{d \Omega_m}{d n_f} - 
    \text{Tr} \frac{d \mathbf{A}^{\text{asym}}}{d n_f}\text{.}
\end{equation}
The full derivative with respect to the occupation number consists of three parts\cite{suIntegrationApproachSecondOrder2015,liChemicalPotentialsOneElectron2024}
\begin{equation} \label{eq:full_derivative}
    \frac{d}{d n_f} = 
    \frac{\partial }{\partial n_f} + 
    \sum_p \frac{\partial }{\partial \epsilon_p} \frac{d \epsilon_p}{d n_f} +
        \left \{ 
        \sum_p \frac{\delta}{\delta \psi_p } \frac{d \psi_p }{d n_f} 
        + c.c \right \}\text{.}
\end{equation}
In Eq.~\ref{eq:full_derivative},
the first term is dominant and corresponds to the explicit dependence on the occupation number,
the second and third term correspond to the orbital relaxation through orbital energies and orbitals.

To evaluate the derivative of the excitation energy $\Omega_m$,
the excitation energy is expressed in terms of the left and the right eigenvectors of the asymmetric RPA matrix 
\begin{equation}
    \Omega_m = (\mathbf{L}^m)^T \mathbf{M} \mathbf{R}^m\text{,}
\end{equation}
where the right eigenvector is $\mathbf{R}^m = \begin{pmatrix} \mathbf{X}^m \\ \mathbf{Y}^m \end{pmatrix}$, 
the left eigenvector is $\mathbf{L}$ obtained from $\mathbf{L}^T = \mathbf{R}^{-1}$ and the RPA matrix in Eq.\ref{eq:rpa_asym} is
$\mathbf{M} = 
\begin{pmatrix} 
    \mathbf{A}^{\text{asym}}  & \mathbf{B}^{\text{asym}} \\ 
    -\mathbf{B}^{\text{asym}} & -\mathbf{A}^{\text{asym}}
\end{pmatrix}$.
Section.2 in the Supplemental Material shows that the right eigenvector $\mathbf{R}^{m}$ is equivalent to the eigenvector in the symmetric problem apart from normalization and the derivative of the excitation energy $\Omega_m$ to the occupation number $n_f$ is
\begin{equation}\label{eq:excitation_energy_derivative}
    \frac{d \Omega_m}{d n_f} 
    = \frac{d [ (\mathbf{L}^m)^T \mathbf{M} \mathbf{R}^m ]}{d n_f} 
    = (\mathbf{L}^m)^T \frac{d \mathbf{M}}{d n_f} \mathbf{R}^m \text{.}
\end{equation}
The detailed expressions for derivative of $\mathbf{A}$ and $\mathbf{B}$ matrices to the occupation number are given in Section.2 in the Supplemental Material. 
This completes our first approach to the RPA chemical potentials.

Next, 
we consider the the RPA correlation energy as the functional of $G_s$,
which is the correlation energy in the Klein $GW$ functional in MBPT\cite{kleinPerturbationTheoryInfinite1961,martinInteractingElectrons2016}.
We also applied the functional derivative approach\cite{liChemicalPotentialsOneElectron2024} to calculate the chemical potential.
By defining the special trace
\begin{equation}
    \mathfrak{Tr} = \int^\infty_{-\infty} \frac{\text{d}\omega}{2\pi i} e^{i\omega \eta} \text{Tr} \text{,}
\end{equation}
the RPA chemical potential can be evaluated via the chain rule of $G_s$
\begin{equation}\label{eq:chain_rule}
    \mu^{\text{RPA}}_{\text{c}} 
    = \frac{d E^{\text{RPA}}_{\text{c}}}{d n_f} 
    =  \mathfrak{Tr} \left [
    \frac{\delta E^{\text{RPA}}_{\text{c}}}{\delta G_s(\omega)} 
    \frac{d G_s(\omega) }{d n_f} \right ]\text{,} 
\end{equation}
where the fractional $G_s$ is defined as\cite{yangExtensionManybodyTheory2013}.
\begin{equation} \label{eq:green_function}
    G_s(x_1, x_2, \omega) = 
    \sum_i \frac{n_i \psi^*_i (x_1) \psi_i (x_2)}{\omega - \epsilon_i - i\eta}  
    + \sum_a \frac{(1-n_a) \psi^*_a (x_1) \psi_a (x_2)}{\omega - \epsilon_a + i\eta}, 
\end{equation}
In common MBPT calculations (for integer electron systems),
the following $GW$ self-energy has been used as the functional derivative of the RPA correlation energy to the non-interaction Green's function\cite{martinInteractingElectrons2016,rambergerAnalyticInteratomicForces2017,vooraEffectiveOneparticleEnergies2019,vooraVariationalGeneralizedKohnSham2019} in Eq.~\ref{eq:chain_rule}
\begin{equation}\label{eq:derivative_energy_to_g}
     \frac{\delta E^{\text{RPA}}_{\text{c}}}{\delta G^N_s(\omega)}=
     \frac{1}{2 \pi i}
     \Sigma^{GW}_{\text{c}} (\omega,0)  \text{.}
\end{equation}
However, for chemical potential calculations, 
we have to carefully consider fractional frontier orbital occupations and its limits approaching the corresponding integer system.
With the details given in SI,
we obtain the RPA self-energy for systems with fractional orbital occupations as
\begin{equation}\label{eq:rpa_self_energy}
    \frac{\delta E^{\text{RPA}}_{\text{c}}}{\delta [G^{N+\delta}_s]_{qp}(\omega)} = 
    [\Sigma^{GW}_{\text{c}}]_{pq} (\omega,\delta) = \sum_{ia,m}
        \left [
        \frac{n_i (pi|\rho_m) (\rho_m|iq)}{\omega - \epsilon_i + \Omega_m - i\eta}
        + \frac{(1-n_a) (pa|\rho_m) (\rho_m|aq)}{\omega - \epsilon_a - \Omega_m + i\eta}
    \right ] \text{,}
\end{equation}
where the transition density is constructed with eigenvectors of the symmetric RPA matrix $\rho_m (x) = \sum_{ia} (\tilde{X}^m_{ia} + \tilde{Y}^m_{ia}) \sqrt{n_i (1-n_a)} \psi_i^* (x) \psi_a(x)$.
The expression of the RPA self-energy in Eq.\ref{eq:rpa_self_energy} looks formally the same for systems with both fractionals $\delta>0$ and $\delta<0$. 
However,
the summation with excitation energies $m$ involved many different and additional terms, 
because the RPA eigenvalue problems in Eq.\ref{eq:rpa_asym} are different depending on the sign of the fractional $\delta$. 
Compared to the integer expression, 
Eq.\ref{eq:derivative_energy_to_g}, for $\delta<0$, there are additional $N_\text{occ}$ excitation energies $\Omega_{m}=\epsilon_\text{HOMO}-\epsilon_{i}, i=1,...N_\text{occ}$; 
for $\delta>0$, there are additional $N_\text{vir}$ excitation energies, 
$\Omega_{m}=\epsilon_{a}-\epsilon_\text{LUMO}, a=1,...N_\text{vir}$.

Therefore, we see clearly the self-energy $\Sigma_\text{c}^{GW}(\omega,\delta)$ is discontinuous, 
as the fractional $\delta$  approaches zero, 
it has two limits: 
$\Sigma_\text{c}^{GW}(\omega,-)$ and $\Sigma_\text{c}^{GW}(\omega,+)$ from both sides. 
Furthermore, neither are equal to the commonly used GW self-energy, 
$\Sigma_\text{c}^{GW}(\omega,0)$ in Eq.\ref{eq:derivative_energy_to_g}, 
long assumed to the functional derivative. 
The correct discontinuous self-energy limits $\Sigma_\text{c}^{GW}(\omega,-)$ or $\Sigma_\text{c}^{GW}(\omega,+)$, 
leads to the corresponding correct RPA chemical potentials as will be shown numerically, 
which $\Sigma_\text{c}^{GW}(\omega,0)$ fails to do. 
Therefore,
the RPA self-energy is defined up to a constant when the Green's function conserves the particle number.
As shown in Ref.\citenum{liChemicalPotentialsOneElectron2024}, 
the functional derivative approach has also been applied to calculate the derivative of the MP2 correlation energy with respect to the occupation number. 
Complete detailed expressions for the RPA chemical potentials from both direct and functional derivative approaches are provided in Section 2 of the Supplementary Material.

Now we compare chemical potentials obtained from two approaches with the numerical finite difference derivatives.
The correlation parts of the derivative of RPA correlation energy with respect to the HOMO and LUMO occupation number of molecular systems are tabulated in Table.\ref{tab:rpa_correlaton_mu_homo} and Table.\ref{tab:rpa_correlaton_mu_lumo}.
It can been seen that both direct derivative and functional derivative approaches provide excellent agreements with the finite difference approach.
The errors of calculating the derivative with respect to the HOMO and LUMO occupation number are only around 0.01 eV when including the orbital relaxation effect by solving the coupled-perturbed Hartree-Fock or GKS equation\cite{popleDerivativeStudiesHartreefock1979,frischDirectMP2Gradient1990}.
In the functional derivative approach,
the RPA self-energy of the ($N+\delta$)-electron system is constructed to obtain the correct derivative of the RPA correlation energy with respect to the HOMO and the LUMO occupation number.
However,
using the RPA self-energy $\Sigma_\text{c}^{GW}(\omega,0)$ evaluated at the $N$-electron system fails to provide the correct derivative of the RPA correlation energy to the occupation number.
As indicated in the last column of Table.\ref{tab:rpa_correlaton_mu_homo} and Table.\ref{tab:rpa_correlaton_mu_lumo},
derivatives obtained from the functional derivative approach using $\Sigma_\text{c}^{GW}(\omega,0)$ provides much smaller values than the finite difference results.

The $GW$ self-energy evaluated for the integer system is commonly used in the $GW$ approximation to calculate quasiparticle energies for predicting accurate IPs and EAs.
However,
the $GW$ quasiparticle energy is not the RPA chemical potential,
which suffers from the massive delocalization error in RPA as first shown in Refs.\citenum{mori-sanchezFailureRandomPhase2009,mori-sanchezFailureRandomphaseapproximationCorrelation2012}.
Unlike the results of using the functional derivative approach for the MP2 correlation energy\cite{liChemicalPotentialsOneElectron2024},
the $GW$ self-energy evaluated at the integer electron system (or constrained to integer systems) can not be used for left or right derivatives.
Despite that the RPA correlation energy can be expressed as an explicit functional of $G_s$,
our results show that the functional derivative of the RPA correlation energy to $G_s$ does not exist for physical systems with integer electron numbers.
Therefore,
our theoretical derivation and numerical results demonstrate that the RPA correlation energy has a significant discontinuity with respect to $G_s$,
which is equal to $\Sigma_\text{c}^{GW}(\omega,+)$ and $\Sigma_\text{c}^{GW}(\omega,-)$ when passing through the integer electron numbers.

\begin{table}[ht]
\setlength\tabcolsep{12pt}
\centering
\caption{The derivative of RPA correlation energy with respect to the HOMO occupation number obtained from different approaches.
The difference of the electron number was $10^{-4}$ in the finite difference approach.
Functional derivative ($\Sigma_\text{c}^{GW}(\omega,0)$) means that the self-energy is evaluated at integer electron numbers.
Geometries were taken from Ref.\citenum{vansettenGW100BenchmarkingG0W02015} and def2-SVP basis set was used.
MSE and MAE stand for mean signed error and mean absolute error.
All values in eV.
The corresponding complete chemical potential comparison are presented in SI.}
\label{tab:rpa_correlaton_mu_homo}
\begin{tabular}{ccccc}
\toprule
           & finite diff & direct derivative & functional derivative & functional derivative ($\Sigma_\text{c}^{GW}(\omega,0)$) \\
\midrule
\ce{B2H6}  & -4.65  & -4.64  & -4.64  & -0.55 \\
\ce{C2H4O} & -6.62  & -6.60  & -6.60  & -1.29 \\
\ce{C2H4}  & -5.08  & -5.08  & -5.07  & 0.01  \\
\ce{C2H6O} & -6.50  & -6.48  & -6.48  & -1.23 \\
\ce{C3H6}  & -5.36  & -5.35  & -5.34  & -0.64 \\
\ce{CH2O2} & -7.50  & -7.49  & -7.49  & -1.44 \\
\ce{CH4O}  & -7.05  & -7.04  & -7.04  & -1.24 \\
\ce{CH2O}  & -7.00  & -6.98  & -6.98  & -1.18 \\
\ce{N2H4}  & -6.45  & -6.44  & -6.44  & -1.08 \\
\ce{H2O}   & -9.47  & -9.46  & -9.47  & -1.35 \\
\ce{H2}    & -8.02  & -8.01  & -8.01  & -0.06 \\
\ce{N3H}   & -5.94  & -5.94  & -5.94  & -0.69 \\
\ce{H2O2}  & -8.09  & -8.07  & -8.07  & -1.75 \\
\ce{Li2}   & -2.65  & -2.65  & -2.65  & 0.05  \\
\ce{NH3}   & -8.02  & -8.01  & -8.02  & -1.01 \\
\ce{H2S}   & -5.96  & -5.95  & -5.95  & -0.41 \\
MSE        &        & -0.01  & -0.01  & -5.66 \\
MAE        &        & 0.01   & 0.01   & 5.66  \\
\bottomrule
\end{tabular}
\end{table}

\begin{table}
\setlength\tabcolsep{12pt}
\centering
\caption{The derivative of RPA correlation energy with respect to the LUMO occupation number obtained from different approaches.
The difference of the electron number was $10^{-4}$ in the finite difference approach.
Functional derivative ($\Sigma_\text{c}^{GW}(\omega,0)$) means that the self-energy is evaluated at integer electron numbers.
Geometries were taken from Ref.\citenum{vansettenGW100BenchmarkingG0W02015} and def2-SVP basis set was used.
MSE and MAE stand for mean signed error and mean absolute error.
All values in eV.
The corresponding complete chemical potential comparison are presented in SI.}
\label{tab:rpa_correlaton_mu_lumo}
\begin{tabular}{ccccc}
\toprule
           & finite diff & direct derivative & functional derivative & functional derivative ($\Sigma_\text{c}^{GW}(\omega,0)$) \\
\midrule
\ce{B2H6}  & 4.56   & 4.56   & 4.56   & 0.99 \\
\ce{C2H4O} & 5.86   & 5.86   & 5.86   & 1.24 \\
\ce{C2H4}  & 5.41   & 5.41   & 5.41   & 1.00 \\
\ce{C2H6O} & 3.42   & 3.41   & 3.41   & 0.57 \\
\ce{C3H6}  & 3.34   & 3.34   & 3.34   & 0.61 \\
\ce{CH2O2} & 6.20   & 6.20   & 6.20   & 1.11 \\
\ce{CH4O}  & 3.57   & 3.56   & 3.57   & 0.48 \\
\ce{CH2O}  & 6.19   & 6.20   & 6.20   & 1.04 \\
\ce{N2H4}  & 3.82   & 3.82   & 3.82   & 0.52 \\
\ce{H2O}   & 4.32   & 4.32   & 4.32   & 0.35 \\
\ce{H2}    & 4.60   & 4.59   & 4.59   & 0.16 \\
\ce{N3H}   & 5.87   & 5.91   & 5.91   & 1.09 \\
\ce{H2O2}  & 4.08   & 4.07   & 4.07   & 0.43 \\
\ce{Li2}   & 2.10   & 2.10   & 2.10   & 0.28 \\
\ce{NH3}   & 3.88   & 3.88   & 3.88   & 0.41 \\
\ce{H2S}   & 4.19   & 4.18   & 4.19   & 0.49 \\
MSE        &        & 0.00   & 0.00   & 3.79 \\
MAE        &        & 0.01   & 0.00   & 3.79 \\
\bottomrule
\end{tabular}
\end{table}

The behavior of functional derivatives of different types of XC functionals in terms of the non-interacting density $\rho_s(x)$\cite{perdewPhysicalContentExact1983,cohenChallengesDensityFunctional2012}, 
the non-interacting density matrix $\gamma_{s}(x,x')$\cite{mori-sanchezDiscontinuousNatureExchangeCorrelation2009} and the non-interacting $G_s(x,x',\omega)$ (from present work) are summarized in Table.\ref{tab:discontinuity}.
The simplest approximation LDA and GGA functionals have continuous derivatives in terms of $\rho_s$.
Because $\rho_s$ can be obtained from the diagonal elements of $\gamma_s$ and $\gamma_s$ can be obtained from $G_s$,
LDA and GGA functionals also have continuous derivatives in terms of $\gamma_s$ and $G_s$.
Similarly,
hybrid functionals have continuous derivative in terms of $\gamma_s$ and therefore in terms of $G_s$.
As shown in Ref.\citenum{liChemicalPotentialsOneElectron2024},
the MP2 chemical potential can be obtained from the functional derivative approach.
Thus,
the functional derivative of the MP2 correlation energy with respect to $G_s$ is continuous.
In DFT, for the exact functional, the derivative discontinuity with respect to the density\cite{perdewPhysicalContentExact1983,shamDensityFunctionalTheoryEnergy1983},
or with respect to the density matrix\cite{mori-sanchezDiscontinuousNatureExchangeCorrelation2009} is well known.

In this work, the discontinuity in the derivative of the RPA correlation energy with respect to $G_s$ is demonstrated. 
This derivative discontinuity can be equivalently viewed: 
the GW correlation energy is discontinuous with respect to the variation of the Green's function, 
when evaluated at $G_s$. 
Because $\rho_s$ or $\gamma_s$ determines $G_s$, 
the discontinuity in the derivative of the RPA energy with respect to $G_s$ implies a corresponding discontinuity in its derivative with respect to $\rho_s$ and $\gamma_s$. 
Similar to RPA, 
the exact XC functional is an infinite summation of diagrams of $G_s$. 
We expect that the exact XC functional is likely to have functional derivative discontinuity with respect to Green's function.

When there is no discontinuity, as in the case of MP2, 
we can derived a non-interacting reference one-electron Hamiltonian based on the energy functional derivative with respect to the $G_s$ and then the functional derivative of $G_s$ with respect to the density matrix $\gamma_s$\cite{liChemicalPotentialsOneElectron2024}. 
In the case of RPA correlation energy, 
because of the presence of the derivative discontinuity, 
the same chain rule leads to two one-electron Hamiltonians, 
one from $\Sigma_{\text{c}}^{GW}(\omega,-)$ for the occupied states and another from $\Sigma_{\text{c}}^{GW}(\omega,+)$ for unoccupied states. 
In order words, 
the two Hamiltonians is a manifestation of RPA correlation energy derivative discontinuity with respect to $\gamma_s$.

Consequently, the RPA fundamental gap, IP-EA, 
comes from the difference of two eigenvalues, 
not of the same Hamiltonian, 
but each of a different Hamiltonian. 
If one insists on using the HOMO-LUMO gap of one Hamiltonian, then the fundamental gap as predicted by $\frac{\partial E}{\partial N}\bigg|_{+}-\frac{\partial E}{\partial N}\bigg|_{-}$, 
the discontinuity of the chemical potential, is the HOMO-LUMO gap, $\Delta\epsilon^{\text{GKS}}$ plus a derivative discontinuity $\mathcal{G}_{\text{xc}}$, 
the values of which depends which one-electron Hamiltonian is used. 
We can capture this in the following equation:
\begin{equation}
    \frac{\partial E}{\partial N}\bigg|_{+}-\frac{\partial E}{\partial N}\bigg|_{-}=\Delta\epsilon^{\text{GKS}}+\mathcal{G}_{\text{xc}} \text{.}
\end{equation}
We note that for the exact functional, 
there is a similar derivative discontinuity constant for the difference between fundamental gap and the KS HOMO-LUMO gap\cite{perdewPhysicalContentExact1983,shamDensityFunctionalTheoryEnergy1983}, 
and also for the difference between fundamental gap and the GKS HOMO-LUMO gap\cite{mori-sanchezDiscontinuousNatureExchangeCorrelation2009} for strongly correlated systems.

\begin{table}
\setlength\tabcolsep{5pt}
\centering
\caption{\label{tab:discontinuity}Summary of functional derivative discontinuity in different types of XC functionals in terms of the density, density matrix and $G_s$.
Check mark \checkmark means derivative is continuous and cross mark $\times$ means discontinuous.
The exact XC functional is expected to have derivative discontinuity in terms of $G_s$ as indicated by the ? mark.
}
\begin{tabular}{c|ccccc}
\toprule
                   & LDA/GGA    & hybrid     & MP2        & RPA      & exact     \\
\midrule
$\rho_s(x)$        & \checkmark & $\times$   & $\times$   & $\times$ & $\times$  \\
$\gamma_s(x,x')$   & \checkmark & \checkmark & \checkmark & $\times$ & $\times$  \\
$G_s(x,x',\omega)$ & \checkmark & \checkmark & \checkmark & $\times$ & $\times$? \\
\bottomrule
\end{tabular}
\end{table}

\begin{figure}[ht]
\includegraphics[width=0.4\textwidth]{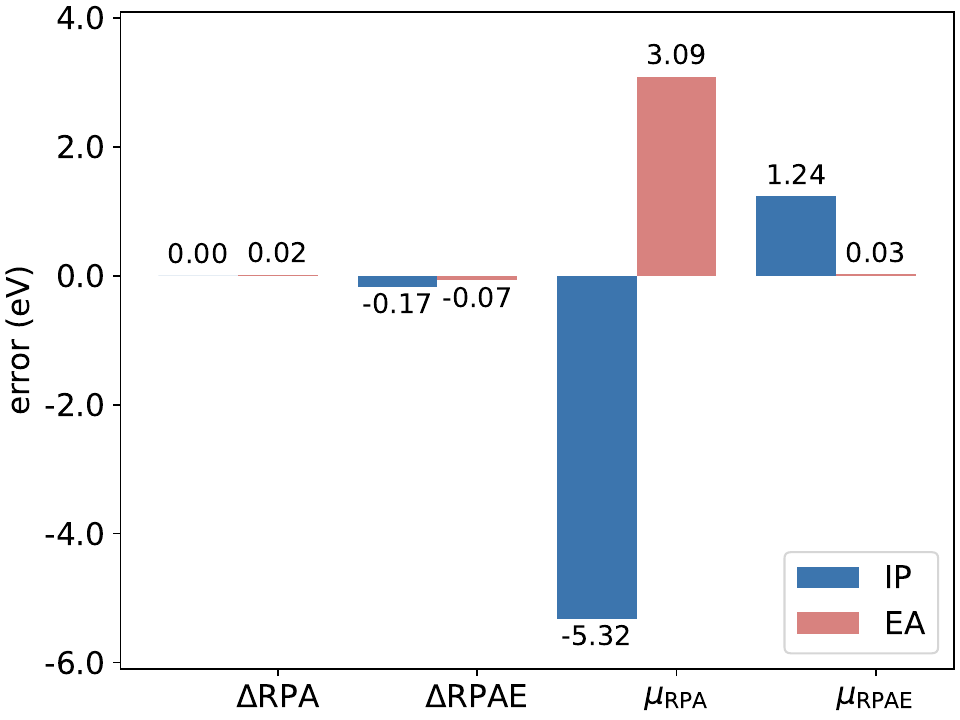}
\caption{Mean signed errors of different methods for predicting ionization potentials and electron affinities of molecules compared to $\Delta$CCSD(T) references.
$\mu_\text{RPA}$ and $\mu_\text{RPAE}$ results were obtained from the finite difference approach for chemical potentials.} 
\label{fig:ip_ea_error}
\end{figure}

Finally,
we examine the performance of the RPA chemical potential for predicting IPs and EAs.
RPA with exchange (RPAE) that has smaller delocalization error than RPA\cite{mori-sanchezFailureRandomPhase2009,mori-sanchezFailureRandomphaseapproximationCorrelation2012} is included for the comparison. 
Mean absolute errors (MAEs) of using HF, $\Delta$RPA, $\Delta$RPAE, RPA and RPAE for predicting IPs and EAs of molecular systems are shown in Fig.\ref{fig:ip_ea_error}.
RPA and RPAE chemical potentials were obtained from the finite difference approach.
$\Delta$CCSD(T) results were used as the references.
Because both RPA and RPAE provide good description for describing systems with an integer electron number,
$\Delta$-based approaches $\Delta$-RPA and $\Delta$-RPAE predict accurate IPs and EAs with errors around 0.3 eV.
As shown in Refs.\citenum{mori-sanchezFailureRandomphaseapproximationCorrelation2012},
RPA suffers from the massive delocalization error.
The correct RPA chemical potential gives huge errors that exceed 3 eV.
Since RPAE has much smaller delocalization error compared to RPA,
the RPAE chemical potential significantly outperforms the RPA chemical potential for predicting IPs and EAs with errors around 0.5 eV,
which highlights the importance of minimizing the delocalization error in constructing exchange-correlation functionals.

\FloatBarrier

In summary,
we present two approaches for calculating the RPA chemical potentials.
The direct derivative approach explicitly takes the derivative of the RPA correlation energy with respect to the electron number,
and the functional derivative approach utilizing the chain rule through the non-interacting Green's function.
Both approaches provide well agreement with finite difference derivatives.
In the functional derivative approach,
the RPA chemical potential is obtained by evaluating the $GW$ self-energy for systems with a fractional electron number.
Our work shows that the functional derivative of the RPA or the $GW$ correlation energy with respect to the Green's function is discontinuous at the non-interacting Green's function with integer electron numbers. 
The exact exchange-correlation functional is expected to have similar derivative discontinuity,
which is analogous to the proven derivative discontinuity with respect to the density and density matrix.
We also showed that the RPAE chemical potentials outperform the RPA chemical potentials for predicting IPs and EAs for molecular systems,
which emphasizes the importance of minimizing delocalization error in developing approximate XC functionals. 
Establishing the derivative discontinuity in RPA correlation energy in the present work should play an important role in future understanding and development of correlation energy functionals through MBPT.

\FloatBarrier

\begin{acknowledgments}
J. L. acknowledges Christopher Hillenbrand at Yale University for the help in the numerical implementation.
W.Y. acknowledges the support from the National Institutes of Health (1R35GM158181).
\end{acknowledgments}

\bibliography{ref}

\end{document}